# The prices of renewable commodities: A robust stationarity analysis

**Manuel Landajo**[1]

**Department of Applied Economics**

**University of Oviedo**

**Avenida del Cristo, s/nº**

**33011 Oviedo**

**Asturias (Spain)**

**Phone: +34 985105055**

**Email: landajo@uniovi.es**

**María José Presno**

**Department of Applied Economics**

**University of Oviedo**

**Avenida del Cristo, s/nº**

**33011 Oviedo**

**Asturias (Spain)**

**Phone: +34 985103758**

**Email: mpresno@uniovi.es**

## Abstract

This paper addresses the problem of testing for persistence in the effects of the shocks affecting the prices of renewable commodities, which have potential implications on stabilization policies and economic forecasting, among other areas. A robust methodology is employed that enables the determination of the potential presence and number of instant/gradual structural changes in the series, stationarity testing conditional on the number of changes detected, and the detection of change points. This procedure is applied to the annual real prices of eighteen renewable commodities over the period of 1900–2018. Results indicate that most of the series display non-linear features, including quadratic patterns and regime transitions that often coincide with well-known political and economic episodes. The conclusions of stationarity testing suggest that roughly half of the series are integrated. Stationarity fails to be rejected for grains, whereas most livestock and textile commodities do reject stationarity. Evidence is mixed in all soft commodities and tropical crops, where stationarity can be rejected in approximately half of the cases. The implication would be that for these commodities, stabilization schemes would not be recommended.

[1] Corresponding author.

## 1. Introduction

The study of the persistence properties of commodity prices is a relevant topic for both theory and econometric reasons. Economically, persistence connects with the weak version of the efficient market hypothesis, implying that future prices cannot be predicted on the past. The distinction between trend-stationary and non-stationary (integrated) processes is also important from the standpoint of stabilization policies because, as pointed out by Reinhart and Wickham (1994), stabilization and hedging strategies are more effective when used against temporary shocks (such as those occurring in stationary, short memory processes), whereas structural policies are required for those having permanent (or temporary but widespread and highly persistent) characteristics. The nature of the shocks in commodity prices is also relevant in forecasting (Diebold and Kilian, 2000) and portfolio risk management (Wang and Tomek, 2007). Ghoshray (2019) enumerates additional reasons stressing the importance of discerning the nature of shocks in the specific case of agricultural commodities. Knowledge of the stochastic properties of the time series also allows for more accurate estimates of the dates of regime transitions, such as those frequently observed in the evolution of commodity prices. Those changes can often be readily connected to specific economic/historic events.[2]

In this paper, we shall focus on the study of the long-run dynamics of the real prices of renewable commodities. A robust methodology, specifically adapted to testing for stationarity in non-linear models, will be employed to analyse the real prices of eighteen renewable commodities in the 1900–2018 study period.

[2] Kilian (2009) points out that some of those events may be included in one of the three categories of supply shocks, aggregate demand shocks, and specific demand shocks. Additionally, Stuermer (2018) and Jacks and Stuermer (2020) characterize positive commodity-specific demand shocks as having immediate, large, and persistent positive effects on real prices.

Theoretical studies on the dynamic stability of equilibria in commodity markets suggest that prices should exhibit some sort of stationarity. For instance, the original Prebisch–Singer hypothesis (Prebisch, 1950; Singer, 1950) assumes that the relative prices of primary commodities in terms of manufactures are stationary around a downward secular trend, and authors like Sapsford (1985), Grilli and Yang (1988), and Ardeni and Wright (1992) advocate for commodity prices to follow that kind of pattern. Along the same line, Deaton and Laroque (1992, 1996) conclude that the random walk hypothesis is implausible because it requires price fluctuations to be permanent. Wang and Tomek (2007) arrive at the same result for the case of agricultural prices, with stationarity being a consequence of the biological nature of commodity production and the cost of arbitrage. The same conclusion is also reached by Williams and Wright (1991) and Peterson and Tomek (2005).

By contrast, Cuddington and Urzua (1989), Cuddington (1992), Kim *et al*. (2003), and Newbold *et al.* (2005) recognize that relative commodity prices may show unit root behaviour. In this regard, recent empirical studies (Gutierrez *et al*., 2015; Baffes and Haniotis, 2016; Dillon and Barrett, 2017; Ghoshray, 2019, among others) have concluded that the shocks to commodity prices are mainly permanent in nature. Meanwhile, Newbold and Vougas (1996) find insufficient evidence of both trend stationarity and difference stationarity. Ghoshray (2019) explains this seeming contradiction between theory and empirical results in terms of factors like prices being denominated in US dollars (thus inheriting the non-stationary behaviour of the US nominal exchange rate; Chen *et al*. 2014), the increasing connection between agricultural and energy prices (Wu *et al*., 2011; Nicola *et al*., 2016), and the effects of monetary policy on commodity prices.

An additional factor behind this contradiction is that the statistical tools may be inadequate. Regarding this, Wang and Tomek (2007) warn about the sensitivity of results to model specification. The time series of commodity prices typically exhibit both high volatility and complex non-linear patterns, including cyclical behaviour and regime changes. The statistical tools have evolved to adapt to that reality. Initially, the empirical works on commodity prices applied unit root tests, typically around a simple linear trend. Subsequent studies (e.g. Leon and Soto, 1997; Kellard and Wohar, 2006; Harvey *et al*., 2010; Ghoshray, 2011; Yamada and Yoon, 2014; Ghoshray *et al*., 2014; Nazlioglu, 2014, among others) explicitly allowed for the presence of structural breaks in the commodity price series.

The potential for the presence of non-linear patterns in commodity prices is also well documented. As mentioned above, Deaton and Laroque (1996) show that the impossibility of negative storage gives rise to non-linearity, whereas Holt and Craig (2006) focused on irreversibility in liquidation decisions pertaining to standing stocks of livestock and perennial crops and its non-linear effect on the prices of those goods. Persson and Teräsvirta (2003) also explicitly consider the possibility of a non-linear pattern in the commodity terms of trade. Harvey *et al.* (2011), relying on quadratic-trend models, apply unit root testing to analyse relative commodity prices. Balagtas and Holt (2009) detect significant evidence of non-linearity, with the null hypothesis of a linear unit root model rejected in favour of a STAR-type alternative in nineteen out of a panel of twenty-four commodities, whereas Enders and Holt (2012) allowed for gradual regime shifts in their analysis. This kind of pattern is reasonable in many cases because the reaction of economic agents to shocks tends to exhibit some delay between the moment when events occur and their reaction.

Our analysis has several distinctive features in comparison with prior works. First, we focus on stationarity instead of unit root testing. The latter has been employed in most studies on commodity prices and is the reference approach in the field; however, unit root tests are known to have low power under stationary but highly persistent processes. Instead, we shall rely on a stationarity testing framework: the null hypothesis of stationarity around the deterministic mean of the process will be tested against the alternative that the data generating process includes an integrated process. This kind of test is interesting because in many cases (e.g. in cointegration analysis), it is more suitable to test for the null of stationarity. Lee and Schmidt (1996) proved that the stationarity test of Kwiatkowski *et al.* (1992) is consistent against stationary long-memory alternatives, so it can be employed to distinguish short-memory stationary processes from long-memory ones. Furthermore, stationarity tests provide a useful means to confirm the results from unit root testing, so both approaches may be regarded as complementary.

Second, the deterministic components in our models potentially include linear and quadratic terms as well as breaks and smooth transition changes in both level and slope, so the specifications are flexible enough to incorporate potential non-linear patterns in the commodity price series. We shall rely on the robust approach outlined in Presno *et al.* (2014). That methodology exploits advances in the time series literature (Perron and Yabu, 2009; Kejriwal and Perron, 2010; Harvey *et al.*, 2010) that enable issues like robust detection of the presence and number of structural changes in the series, stationarity testing, and estimation of change locations. Many studies (e.g. Hadri, 2012; Gevorkyan, 2016; Mintz, 1967; Reinhart and Wickham, 1994) have also shown that commodity prices display complex volatility patterns. The stationarity testing framework we employ also exhibits good performance under non-linear/conditional-heteroskedasticity error processes (Landajo and Presno, 2010).

Finally, we focus on real rather than relative commodity prices. A vast majority of prior research has concentrated on the prices of primary commodities relative to those of manufactured goods because this seems to be the natural subject of analysis when testing the Prebisch–Singer hypothesis of a decreasing trend in those prices. In this paper, we are interested specifically in the potentially integrated nature of the random shocks in commodity prices measured in real terms. As commented above, this goal is relevant for both economic and statistical reasons. The US Consumer Price Index (CPI) will be employed as a deflator. Many studies (e.g. Antonovitz and Green, 1990; Deaton and Laroque, 2003; Ghoshray and Perera, 2016; Jacks, 2019) point out this as a sensible choice, although many other possibilities are available, including the Manufactures Unit Value (MUV) Index, producer price indexes (e.g. Ahrens and Sharma, 1997; Enders and Holt, 2012; Presno *et al.*, 2014), and inflation-bias-corrected CPIs (Cuddington, 2010).[3]

The rest of the paper is structured as follows: Section 2 briefly reviews the methodology. The results of the empirical analysis are presented in Section 3. Section 4 summarizes the conclusions.

## 2. Methodology

As commented above, we follow the approach proposed in Presno *et al.* (2014), which builds on a flexible specification that allows for a large variety of patterns (including both linear and quadratic terms as well as structural changes in the level and slope of the series), so the models should be able to capture the combined effects of long-run deterministic trends, gradual changes (more or less cyclical in nature), and abrupt changes caused by factors like unexpected political, economic, and climatic events.

[3] In Cuddington (2010), an inflation-bias-corrected CPI is employed as a deflator. In that (logarithmic) model, the correction amounts to subtracting a linear term from the logarithm of nominal prices. That change is relevant when testing the Prebisch–Singer hypothesis (because the sign of the trend slope may reverse as a result of that correction) but has no effect on stationarity testing because the subtraction of a deterministic linear trend term leaves the stationary/integrated nature of the original series unaffected. Therefore, the same conclusions are obtained in our stationarity tests, which by construction incorporate a linear deterministic component in the models for the logarithm of commodity prices, regardless of the corrected/uncorrected nature of the CPI employed to deflate the series.

We shall rely on the following error component model for the time series (denoted by $y_t$) of each commodity price:

$$y_t = \mu_t + f(t/T, \boldsymbol{\theta}) + \varepsilon_t, \qquad (1)$$

$$\mu_t = \mu_{t-1} + u_t; t = 1, \ldots, T;\ T = 1,2, \ldots$$

where $\{\varepsilon_t\}$ and $\{u_t\}$ are independent zero-mean random error processes with respective variances of $E(\varepsilon_t^2) = \sigma_\varepsilon^2 > 0$ and $E(u_t^2) = \sigma_u^2 \geq 0$; $\{\mu_t\}$ starts with $\mu_0$, which is assumed to be zero. $f(t/T, \boldsymbol{\theta})$ is the mean of the process. This is a deterministic function of time (with $\boldsymbol{\theta}$ being a vector of free parameters).

We employ several specifications that allow for quadratic, linear, and intercept terms, possibly affected by $n$ regime changes that can be either instant or gradual. More specifically, the following two variants are considered:

**Model I** (quadratic, with *n* changes in level):

$$f(t/T, \boldsymbol{\theta}) = \beta_0 + \beta_1 t/T + \beta_2 (t/T)^2 + \textstyle\sum_{j=1}^{n} \delta_j F(t/T, \boldsymbol{\pi}_j) \qquad (2)$$

with $\theta = (\beta_0, \beta_1, \beta_2, \delta_1, \pi_1, \ldots, \delta_n, \pi_n)$. In the case of instant changes (i.e. breaks), the step function $F(t/T, \boldsymbol{\pi}_j) = 1$ if $(t/T) > \lambda_j$, $F(t/T, \boldsymbol{\pi}_j) = 0$ otherwise, with $\boldsymbol{\pi}_j = \lambda_j \in (0,1)$ being the relative timing of the break point, is employed. The other possibility we consider is the smooth transition case, with the logistic curve $F(t/T, \boldsymbol{\pi}_j) = \left[1 + exp\{-\gamma_j (t/T - \lambda_j)\}\right]^{-1}$, where $\boldsymbol{\pi}_j = (\lambda_j, \gamma_j)$, $\lambda_j$ is the relative position of the transition midpoint into the sample, and $\gamma_j$ controls the speed of transition (gradual for small values of $\gamma_j$ and approaching a break as $\gamma_j$ increases).

**Model III** (quadratic, with *n* changes in both level and slope)[4]:

$$f(t/T, \boldsymbol{\theta}) = \beta_0 + \beta_1 t/T + \beta_2 (t/T)^2 + \textstyle\sum_{j=1}^{n} (\delta_j + \eta_j t/T) F(t/T, \boldsymbol{\pi}_j) \qquad (3)$$

with $\theta = (\beta_0, \beta_1, \beta_2, \delta_1, \eta_1, \pi_1, \ldots, \delta_n, \eta_n, \pi_n)$.

[4] We follow the standard naming in the field (models I to III). Model II, which only allows for shifts in slope, is omitted here, although it is implicitly employed because the strategy of Kejriwal and Lopez (2013) applies unrestricted testing (designed to detect a break in the slope while also allowing for the intercept to shift), with the critical values corresponding to Model II.

We also consider linear specifications by employing the above models with the quadratic terms omitted.

The analysis proceeds for each series in the following three steps:

A. First, statistical tests are applied to determine the potential presence and the number of structural breaks in the series. That number must be consistently estimated, regardless of the (stationary/integrated) nature of the process. The methodology derived in Perron and Yabu (2009), Kejriwal and Perron (2010), and Harvey *et al.* (2010) allows for this in the linear trend setting. In the quadratic case, we apply the extension proposed by Presno *et al.* (2014). Thereafter, to detect changes in the level and/or slope of the series, the methodology of Presno *et al.* (2014), which relies on the sequential testing approach of Kejriwal and Lopez (2013),[5] is applied. Because the time series were relatively short in this case, following Kejriwal and Perron (2010), a maximum of only two structural changes was allowed.

B. Then stationarity is tested, conditional on the number of changes estimated in step A. We rely on the KPSS-type approach of Landajo and Presno (2010) and Presno *et al.* (2014) to test for the null of stationarity around a deterministic function, allowing for linear and quadratic components with (instant/gradual) structural changes. In equation (1) above, under the null of stationarity (i.e. $H_0 : \quad q \equiv \frac{\sigma_u^2}{\sigma_\varepsilon^2} = 0$), $y_t$ is stationary around $f(t/T, \boldsymbol{\theta})$ whereas it is integrated under the alternative (namely $H_1 : \quad q > 0$). The Lagrange multiplier stationarity test statistic has the form

$$\hat{S}_T = \hat{\sigma}^{-2} T^{-2} \sum_{t=1}^{T} \left( \sum_{i=1}^{t} e_i \right)^2 \tag{4}$$

[5] The extended version of the testing procedure of Kejriwal and Lopez proceeds in three steps. First, a model with a single structural break is tested under the most general specification (model III) by using the test proposed by Perron and Yabu (2009). In that case, the test statistic is the *Exp* functional of the Wald test (*ExpW*). If the null of no change is rejected—which may happen because of a change in level and/or slope—the unrestricted test is then applied in the second stage. In that test, rejection may occur because of a change in the growth rate. After evidence of the presence of breaks has been obtained, the null of one versus two breaks is tested by employing the Kejriwal and Perron (2010) procedure as extended to the quadratic case by Presno *et al.* (2014). The test statistic for testing this hypothesis can be expressed as $ExpW\,(2/1) = \max_{1 \le i \le 2} \{ExpW^{(i)}\}$, where $ExpW^{(i)}$ is the one-break test statistic for segment *i*. The null of a single change versus the alternative of two is rejected for large values of *ExpW*(2/1). Because of the potentially low power of the test in the presence of multiple breaks (especially when consecutive changes have opposite signs), we also report (see Section 3 below) the results of the one-versus-two break test, regardless of the conclusions of the single break test. Finally, the number of breaks in the level is also analysed, conditionally on a stable slope obtained in the first step, by employing the test derived by Harvey *et al.* (2010).

with $\{e_i\}$ being the residuals of non-linear least squares fitting of the model and $\hat{\sigma}^2$ being a suitable (e.g. a nonparametric, spectral-window) estimator for the long-run variance of $\{\varepsilon_t\}$. Critical values for the test are calculated by using the Monte Carlo-based bootstrap algorithm proposed by Landajo and Presno (2010).

C. Estimation of change location, conditional on the results of the stationarity testing. Following Kejriwal and Lopez (2013), under the unit root alternative, the change dates are more accurately estimated by fitting model (1) in first differences, whereas under the null of stationarity, more efficient estimates are obtained from the in-level version of the model.

All the above routines were implemented in MATLAB (codes are available from the authors upon request). Further technical details on all the above estimation/testing procedures are provided in Presno *et al*. (2014; 2018).

## 3. Empirical analysis

### *The dataset*

The data to be analysed are the logarithms of the real price indices (annual, employing 1900 as the base year) of eighteen renewable commodities for the 1900–2018 study period. As commented above, the US CPI was used as a deflator. The commodities are those included in the classical study by Grilli and Yang (1988), namely coffee, cocoa, tea, rice, wheat, maize, sugar, beef, lamb, banana, palm oil, cotton, jute, wool, hides, tobacco, rubber, and timber. The price data come from the Grilli–Yang dataset, extended by S. Pfaffenzeller[6] until year 2011. For 2012 to 2018, we updated the series with prices obtained from the Pink Sheet elaborated by the World Bank, excepting the cases of jute (from the Statistical Bulletins of FAO) and sugar (obtained from the International Sugar Organization). Two series, namely wheat and rubber, were discontinued after 2011. For several reasons (primarily heterogeneity and data availability issues), we deemed it more appropriate to avoid extending them after that date.

### *Results*

To begin with, many agricultural commodities have experienced large drops in their real prices throughout the 1900–2018 period, ranging between -41% (jute) and -85% (sugar). The only remarkable exceptions that experienced real increases in this period are livestock (beef, +184%;

[6] Pfaffenzeller *et al*. (2007) provided data up to 2003; thereafter, data up to 2011 are provided by http://www.stephan-pfaffenzeller.com/gycpi/.

lamb, +305%) and tobacco (+73%). Banana (0%) and timber (+10%) essentially remained flat in real terms.[7]

To some extent, that decreasing inertia in real prices would be a mirror image of the rising trend in farm productivity, especially after 1945 (e.g. Wang *et al.*, 2018). In some cases, particularly in non-food commodities, long-run reductions in the real prices can be connected to changes like the progressive displacement during the 20th century of materials like rubber, natural fibres, and hides by cheaper or better performing synthetic substitutes in the production of tires, textiles, and many other products.

It is also well known that this very-long-run evolution has not occurred in a steady way. In many cases, a long-run decreasing evolution has been affected by irregular cyclical patterns and occasional price rebounds connected to specific political, economic, and climatic events.

## A. Detection and number of structural changes

According to the results in Table 1 below, which displays the outcome of the robust procedure for structural change detection, most of the series studied had at least one structural change regardless of the model specification (linear, quadratic) considered. In a large majority of cases, two changes are detected, and only in five series (cocoa, palm oil, wool, hides, and tobacco), the number of changes differs among linear and quadratic models. Additionally, in most cases (with a single exception in both linear and quadratic models), the change occurs in both the slope and level of the series. Finally, in a few series where a contradiction between the *ExpW* and *ExpW*(2/1) tests of Perron and Yabu (2009) and Kejriwal and Perron (2010) arises, we opted for analysing both alternatives in the subsequent study.

--- PLS. INSERT TABLE 1 ABOUT HERE---

## B. Stationarity testing

Given the variety (linear/quadratic; break/smooth transitions) of models for each series, before analysing stationarity, we applied model selection criteria conditional on the number of changes detected in the previous phase. Table 2 below shows that in most of the time series analysed, the complexity penalization criteria (namely Akaike's AIC information criterion, Schwartz's SIC, and modified R-squared) chose models including either breaks or smooth transitions, with a large portion of quadratic models being selected. More specifically, many of them are quadratic specifications with two—either smooth (tea, rice, beef, lamb, banana, jute, rubber, timber) or instant (cocoa, wheat, palm oil, tobacco)—structural changes. In other cases, the deterministic component is linear with breaks (sugar, wool, hides) or with gradual changes (maize, cotton). The case of coffee seems somewhat exceptional, with only a quadratic specification without any changes chosen by the complexity penalty criteria (we adopted the rule of selecting the model that was chosen by all—or, in case of discrepancy, at least two—of the three complexity penalization criteria considered).

[7] Percent variation rates calculated in real terms, comparing the CPI-deflated prices in 1900 (base year) and 2018.

--- PLS. INSERT TABLE 2 ABOUT HERE---

Overall, Tables 1 and 2 strongly suggest the presence of non-linear patterns in a vast majority of the real price series analysed, with quadratic specifications being selected more frequently than linear models, and smooth transition structures chosen in many cases instead of instant breaks.

Thereafter, stationarity testing was carried out for each series under each of the four specifications considered. For the sake of brevity, Table 3 only reports the results for the models selected in the previous phase (those for the other model specifications are available from the authors). The tests suggest that, when allowing for a flexible enough model specification, stationarity around the deterministic part of the series is rejected at 5% significance level in half of the renewable commodities analysed (namely cocoa, beef, lamb, banana, cotton, jute, hides, rubber, and timber), whereas it fails to be rejected for coffee, tea, rice, wheat, maize, sugar, and palm oil.[8] In the cases of wool and tobacco, the evidence is somewhat mixed, with stationarity rejected at 10% but not at 5% significance.

--- PLS. INSERT TABLE 3 ABOUT HERE---

## C. Change estimates

[8] Landajo *et al.* (2021), employing nonparametric stationarity testing, recently analysed the real monthly prices of energy-related commodities for the 1980–2020 period, with stationarity also failing to be rejected in the case of palm oil.

Based on the outcome of the above analysis of the number of changes, model specification, and stationarity, we can fit a model for each series, so the approximate dates and sizes of the structural changes affecting the series can be estimated. A summary of the results is provided below. As noted above, for those series classified as stationary, the models are fitted in levels, whereas when unit roots have been detected, it is more appropriate to fit the models in first differences with a view to obtaining more accurate estimates for the change dates (Kejriwal and Lopez, 2013). The same approach was employed by Presno *et al.* (2014) for non-renewable commodity prices.

The results appear in Table 4 below. Estimates for the exact change dates are displayed for the break models, whereas in the case of gradual changes, the estimates correspond to the transition midpoint. Figure 1 below shows the time series (logarithm of the real price index, continuous line) and its fitted model (broken line) for those series classified as stationary. The fitted models (which estimate the joint effect of the deterministic components of the models) clearly reveal the presence of breaks and smooth transitions in linear and quadratic trends**,** including changes in slope and middle-/long-run oscillations of a more or less cyclical nature.

--- PLS. INSERT TABLE 4 ABOUT HERE---

--- PLS. INSERT FIGURE 1 ABOUT HERE---

As for the changes detected, some of them would be more idiosyncratic in nature, although the effects of broader, well-known historical episodes can also be traced in many cases. Thus, models detect large increases in real commodity prices during the inflationary period of World War I and its immediate aftermath (e.g. in the case of timber, with a change detected around 1918 corresponding to a 46% rise from the previous year). Several models also find structural changes around the depression of 1920–21, a brief episode that came along with sharp drops in the US real GDP (-13% between 1919 and 1921, the steepest decline on record until recently) and US CPI (-16% between 1920 and 1922).[9] These would be the cases of banana (with a gradual change detected by the models around 1920) and palm oil (a break, also in 1920, reflecting a 70% drop from its real price peak in 1918). A change is also detected in 1922 for cotton (after a sharp rebound following the previous crisis) and jute. Overall, the 1920s was a relatively weak and volatile period for many commodities. For sugar, the models detect a break in 1925 (coinciding with a 40% nominal price drop from the previous year), whereas in the case of tobacco, a break is estimated in 1926. As for rubber, the models detect a negative change around 1926–1927. In real terms, rubber prices have been on a decreasing path since 1911 (following the creation of the first synthetic rubbers in 1909–1910), with occasional, sharp rebounds (a 170% yearly increase occurred in 1925, followed by a 30% drop the following year).

[9] Ghoshray *et al*. (2014), citing Hadass and Williamson (2003), also point to the negative roles of a retreat to autarky and the end of low transportation costs after World War I. Cuddington and Urzua (1989), who also detect an abrupt drop in relative commodity prices after 1920, stress the cyclical nature of those shocks.

Not surprisingly, for some commodities, the models also detect structural changes in the 1930s, consistent with the deflationary period of the Great Depression (a time in which large nominal price drops occurred in many renewable commodities, partly offset by a 15% reduction in broad CPI levels across the decade), which reduced the global demand for commodities (e.g. Jacks and Stuermer, 2020; Ocampo and Parra, 2007). This is the case of changes detected by the models in wheat (1930), hides (1930 and 1933, coinciding, respectively, with its maximum and minimum prices of the decade), and rice.

The inflationary period immediately following World War II (with the CPI doubling in the United States between year 1941 and the mid-1950s) is also represented in the models, with several structural changes detected shortly after 1946 (the price ceilings for most products, excepting sugar and rice, were lifted in the United States in November that year; Rosenberg, 2003, p. 50). For instance, cocoa (with a change detected in 1947) rose by 230% in real terms between 1945 and 1948 in that country. Another instance is wool (a change detected in 1949). The case of lamb is exceptional because its real price was steadily decreasing from 1938 to 1950 (an overall drop exceeding 80%), with a stabilization thereafter coinciding with the beginning of the Korean War, followed by a slow reversal from that year on (livestock prices took off after 1955). The models detect a structural change around 1951.

More recent structural changes are also located by the models in the inflationary decade of the 1970s, when sharp rises occurred in many commodity prices. These include cocoa (a change detected in 1973) and tea (in 1970 and 1977, the latter coinciding with a 75% nominal price increase that year). On the contrary, the 1980–2001 period was rather weak for renewable commodities, with many of them trading sideways or in more or less steep downtrends. These include the cases of sugar, with a 50% real drop in 1982 (the models detect a change in 1982–1983) and palm oil (a change detected in 1986, corresponding to a 70% drop from 1984 levels). Structural changes are also found in maize (around year 1988), coinciding with a strong recovery in its nominal price after a weak phase beginning in 1980.

The period between 2002 and the first five months of 2008 was a bull market for many commodities, with the Commodity Research Bureau (CRB) Index growing by more than 250% in nominal terms, although the rises in agricultural products generally were far more modest. Our models estimate structural changes in 2002 for wheat (indeed, a reversal in its real price, after more than 20 years of falling) and wool (a real rise by almost 70% occurred that year), as well as for banana (a change is detected around 2003; the real prices of that fruit bottomed that year, with a 40% drop from 2001 levels, and have been steadily rising since then). A change is also detected for rice in the 2007–2008 period, coinciding with a 90% real price increase.

After the 2008–2009 financial crisis (which cut the value of the CRB Index by half), commodities had a very quick rebound (the models detect a change for cotton in 2011, coinciding with a +130% rally from its minimum in 2009). However, since 2011, most renewable commodity prices were flat or falling, with the only exceptions of two textiles (wool and jute) that experienced mild upswings.

### *Discussion and analysis*

Some of the above structural changes roughly coincide with those reported in previous works, although the deflators and study periods considered in each case generally differ, so some caution is required when comparing results. The structural changes detected for rubber, wool, palm oil, cocoa, and sugar roughly match those found by Kellard and Wohar (2006) for relative commodity prices; those of banana, wool, tobacco, rubber, and maize roughly coincide with Ghoshray *et al*. (2014), and those of sugar and palm oil fit with Ghoshray (2011). The structural change detected around 2007 in the rice series would agree with an upward shift also captured by Enders and Holt (2012).

Many of our models detect structural changes at the beginning of this century, coinciding with a commodity boom (identified around year 2004 by Radetzki, 2006) in that period, pushed up by factors including the explosive growth in the demand of raw materials from China and India and the expansionary monetary response by the Federal Reserve after the attacks of 11 September 2001 and the collapse of the dotcom bubble in the stock markets. Jacks and Stuermer (2020) identify China's rapid industrialization and urbanization as an aggregate commodity demand shock leading to stronger-than-expected increases in the demand for a broad variety of commodities over the past two decades.

Not surprisingly, some of our results can also be roughly interpreted from the standpoint of cycles and super cycles[10] in commodity prices (as commented above, the deterministic part in our models allows for changes and reversals in the mean of the series, so the possibility of cyclical behaviour is implicitly embedded). Erten and Ocampo (2013) address the study of the long-run patterns in commodity prices from that perspective, emphasizing the determining role of expansions in the demand of raw materials and other industrial inputs during the periods of the rapid industrialization and urbanization of various economies (e.g. Europe and Japan post-WWII; the emerging economies of today). The structural changes we detect in the 1920s would fit into the second part (after peaking during World War I) of the first commodity cycle they identify. Those located by our models in the 1930s and the 1945–1955 decade correspond, respectively, to the take-off and the peak of their second cycle, coinciding with the post-war reconstruction of Europe and the economic emergence of Japan. The changes detected in the early 1970s would coincide with the peak of their third commodity cycle, whereas those in the 1980–2000 period would correspond to its strong downward phase. The boom market and changes our models detect in the 2003–2008 period would fit into the early stages of their latest cycle, with the period after 2010 possibly corresponding (in our interpretation) to the end phase of that cycle.

As for the results of stationarity testing, it is observed that half of the series are classified as non-stationary. By groups, we observe that stationarity is rejected for livestock (lamb and beef) and textile commodities (cotton, jute, wool at 10% significance, and hides), whereas it fails to be rejected in the case of grains (rice, wheat, and maize). The evidence seems more mixed in the remaining cases (all of them soft commodities and tropical agricultural products), with stationarity failing to be rejected for coffee, tea, sugar, palm oil, and tobacco (in the latter, at 5% but not at 10% significance), whereas integratedness is obtained for cocoa, banana, rubber, and timber.

---

[10] Jacks (2019) analyses long-run trends, middle-run cycles, and short-run boom/bust episodes in real commodity prices, concluding that commodity price cycles entail large and multi-year deviations from the long-run trends and are punctuated by booms and busts that have large impacts on the commodity-exporting nations. Erten and Ocampo (2013) stress the differences between super cycles and short-term fluctuations: the former tend to span much longer periods (with upswings of 10 to 35 years and complete cycles of 20 to 70 years) and are simultaneously observed in a broad range of commodities. Many recent contributions on middle-term commodity price cycles (e.g. Cuddington and Jerrett, 2008; Erten and Ocampo, 2013; Jerrett and Cuddington, 2008) employ the Christiano and Fitzgerald (2003) band pass filter to detect cycles.

Again, comparisons with previous works require the differences in study periods and deflators to be considered. Kellard and Wohar (2006) apply unit root tests allowing for breaks and conclude that half of the eighteen renewable commodity prices of the Grilli–Yang data set, deflated by the MUV Index, are stationary. They report non-stationarity for banana, beef, cocoa, cotton, and lamb and stationarity for maize, tea, wool, palm oil, and rice, but their conclusions differ from ours for the remaining series. Balagtas and Holt (2009) test linear unit root models versus smooth transition alternatives and find that thirteen out of the same group of relative commodity prices are stationary. The same number of stationary series was detected by Ghoshray *et al.* (2014), employing robust procedures to deal with breaks in the series. Harvey *et al.* (2011) analysed the same series by employing unit root testing (incorporating a local quadratic trend) and concluded that about two thirds of them are stationary around a possibly non-linear trend. Our conclusions differ from theirs in the cases of rubber and timber (stationarity is rejected in our analysis, whereas they reject the unit root hypothesis under both linear and quadratic model specifications) and tea (our tests fail to reject stationarity, whereas theirs do no reject unit roots under any specification). Wang and Tomek (2007) apply unit root tests to monthly agricultural prices in the United States for the 1960–2005 period. Their study includes nominal prices, real prices deflated by the US CPI, and the logarithms of both. They find that stationarity is predominant in the case of nominal prices, with the results being sensitive to the inclusion/omission of structural changes in the models. Ghoshray (2019) recently extended Wang and Tomek’s dataset and employed unit root tests that allow for structural breaks and non-stationary volatility to conclude that only one third of the series they analysed are stationary. Nazlioglu (2014) uses a panel stationarity test and finds evidence in favour of the trend-stationary nature of the prices of twenty-four (both renewable and non-renewable) commodities. Enders and Holt (2012) analysed monthly prices (beginning in 1960 and deflated by the Producer Price Index) and applied both unit root and stationarity tests to a Fourier-based specification that allows for mean shifts, concluding that commodity prices revert to a smoothly evolving mean.

**Policy implications**

The evolution of renewable commodity prices remains a highly relevant matter for both developing and developed nations. In industrial countries, commodity prices affect the inflation rates through the cost of the inputs for manufactured goods and services as well as the national economic growth rates and sectoral and spatial allocations of world capital flows (Cashin *et al*., 2000). In the case of developing nations, the real prices of commodities directly affect both economic growth and poverty level, given the huge weight of raw material exports in the trade basket of most of these countries. Thus, the stationary/integrated nature of commodity prices certainly has a strong effect on the income and consumption levels of those countries because the persistence in commodity price shocks induces large fluctuations in their earnings. Stabilization policies to smooth the income flows are known to be more effective when used against temporary shocks (Reinhart and Wickham, 1994), allowing for external borrowing to balance national income and consumption.[11] According to our results, this would apply in particular to coffee, tea, sugar, palm oil, maize, rice, wheat, and possibly wool and tobacco, from the list of commodities analysed, although (as also pointed out in the literature) the presence of structural changes in the price series certainly may render the application of stabilization programs quite problematic in many cases.

At the other extreme, price shocks are typically long-lived under difference stationarity, with the stabilization programs tending to be hard to implement and their operating costs often exceeding the benefits of consumption/income smoothing. Thus, adjustment to the new long-run levels of these magnitudes would be the preferred policy response (Cashin *et al*., 2000). In our case, given the seemingly integrated nature of the real prices of banana, cocoa, rubber, timber, jute, cotton, beef, lamb, and hides, stabilization policies seem ill-advised for those specific commodities. Indeed, as pointed out by Ghoshray *et al.* (2014), stabilization programs for many commodities (including cocoa, coffee, and jute) have been abandoned since the late 1980s.

## 4. Concluding remarks

In this paper, we have analysed trend stationarity for the annual real prices of a representative set of renewable commodities for the 1900–2018 study period. We employed a robust methodology that enables the detection of structural changes in the price series and their inclusion, along with other non-linear features, in the testing process.

Overall, the evolution of the deflated prices of most renewable commodities has been affected by either instant or gradual changes of regime, which correspond in many cases to well-known political, economic, and natural events, including wars and inflation/deflation periods. Those structural changes have occurred in a general landscape of strong productivity growth in the agricultural sector, specifically following the arrival of artificial fertilizers and an impressive number of other technical innovations following the end of World War II. Technology advances have also enabled a gradual substitution of many non-food agricultural commodities by cheaper or better-performing synthetic analogues.

[11] Kellard and Wohar (2006) point to the finite nature of shocks, their amplitude, and the time required to revert to equilibrium as determining factors for the success of price stabilization policies.

The results of our analysis confirm that most of the price series analysed display patterns that include possibly non-linear long-run trends and all kinds of both gradual and abrupt regime transitions. Once these features are suitably incorporated into the models, stationarity is rejected in half of the series, with some differences observed among groups. Integratedness would be predominant in livestock and textile commodities, whereas all cereals are classified as stationary. As for the remaining price series, stationarity is rejected in roughly half of them.

The above findings for renewable commodity prices deflated by the US CPI are qualitatively similar to recent results (obtained by employing sophisticated unit root testing) for relative (MUV-index-deflated) commodity prices. They also fail to support stabilization policies because these would be ineffective against non-stationary shocks and hardly viable in the case of most of the stationary series analysed, given the implementation issues that arise in the presence of non-linear patterns.

## Conflict of interest

None.

## Data availability statement

The data that support the findings of this study are openly available in Mendeley Data at https://data.mendeley.com/datasets/gtgy2ntkn9/1

**Funding:** The authors acknowledge the financial support from grant PID2020-115183RB-C21, funded by the Spanish Ministry of Science and Innovation (MCIN/ AEI//10.13039/501100011033).

**Acknowledgments:** We wish to thank the Associate Editor and an Anonymous Reviewer for their comments, which have helped to significantly improve this work. Any remaining shortcomings are responsibility of the authors.

# TABLES AND FIGURES

Figure 1: Logarithms of real commodity prices (continuous line) and fitted models (broken line) for the series classified as stationary.

| Coffee | Tea | Rice |
| --- | --- | --- |
| 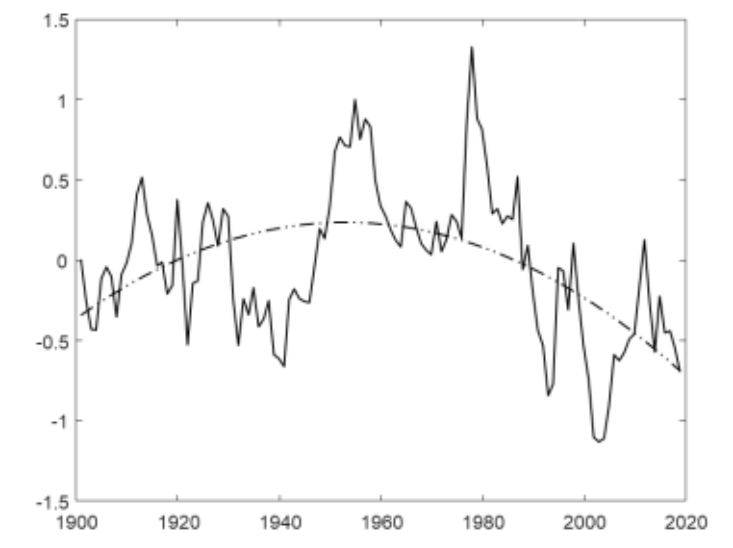 | 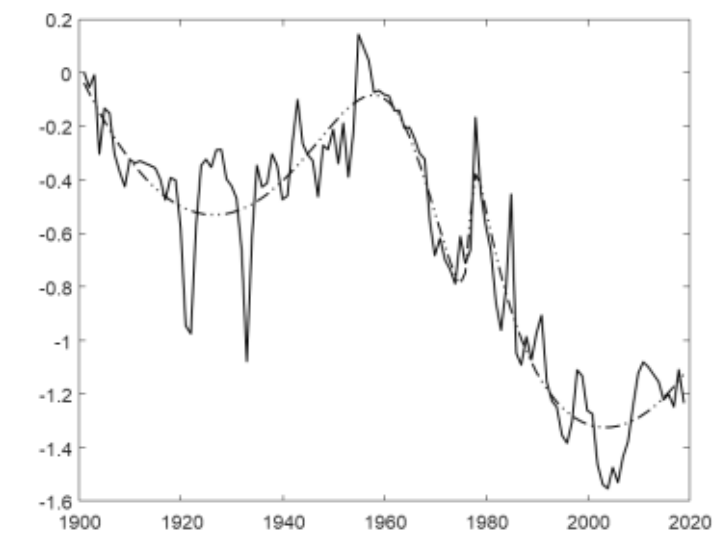 | 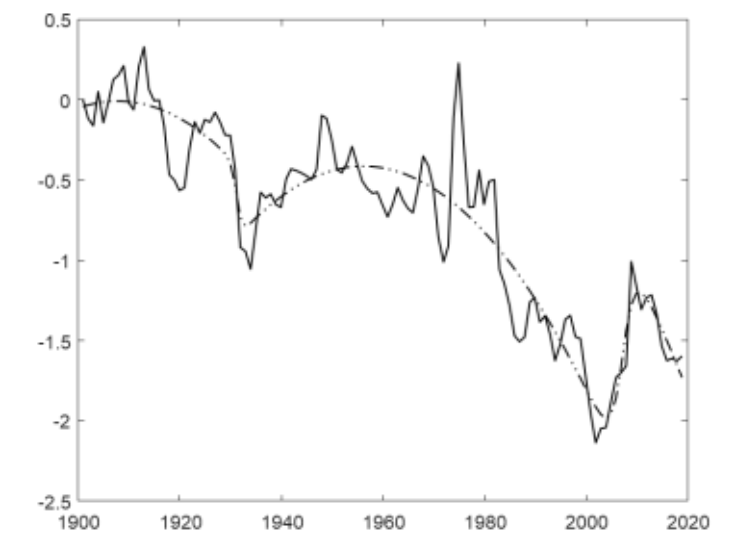 |
| Wheat | Maize | Sugar |
| 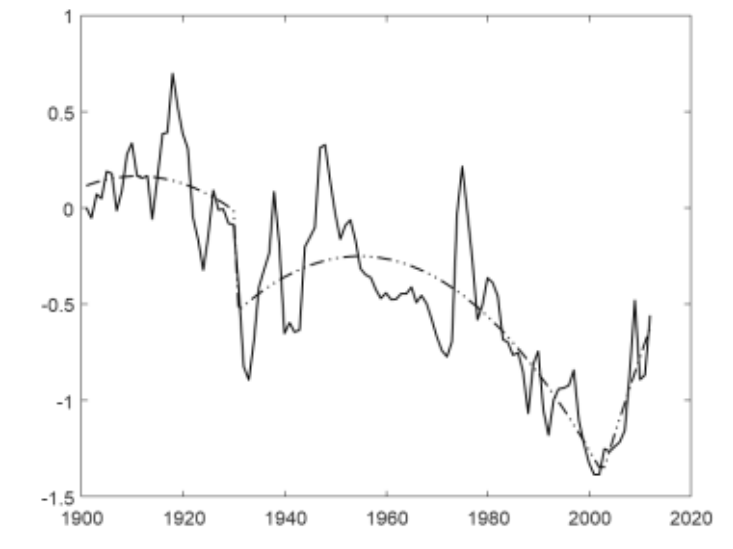 | 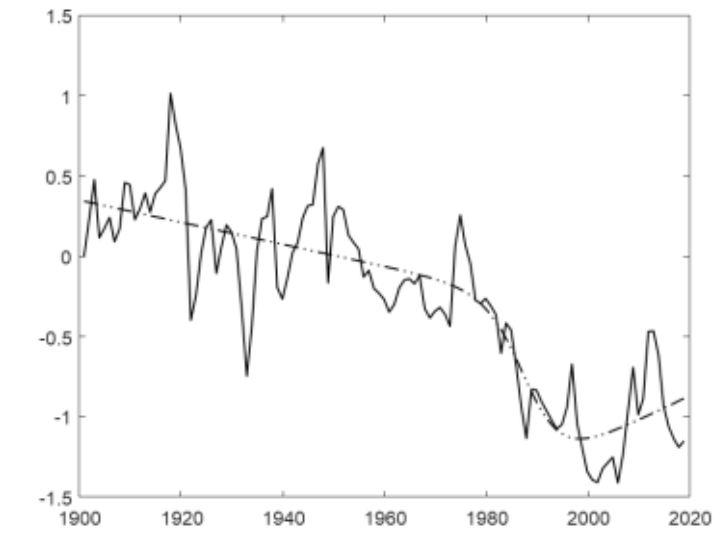 | 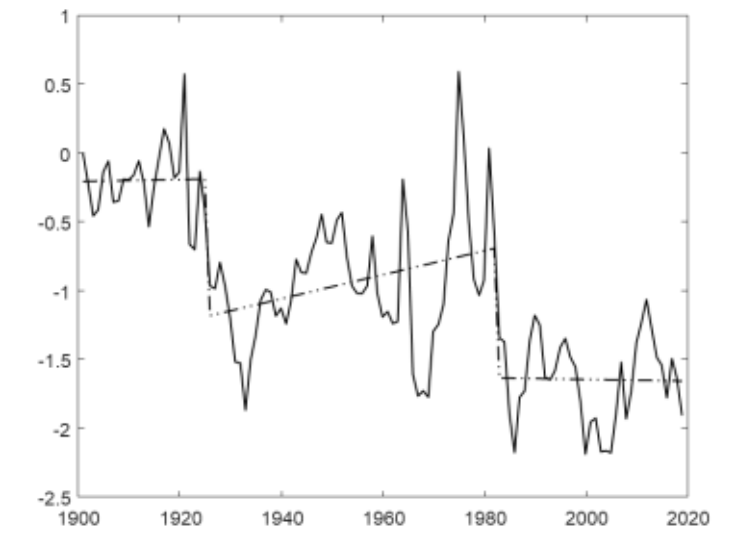 |
| Palm oil | Wool | Tobacco |
| 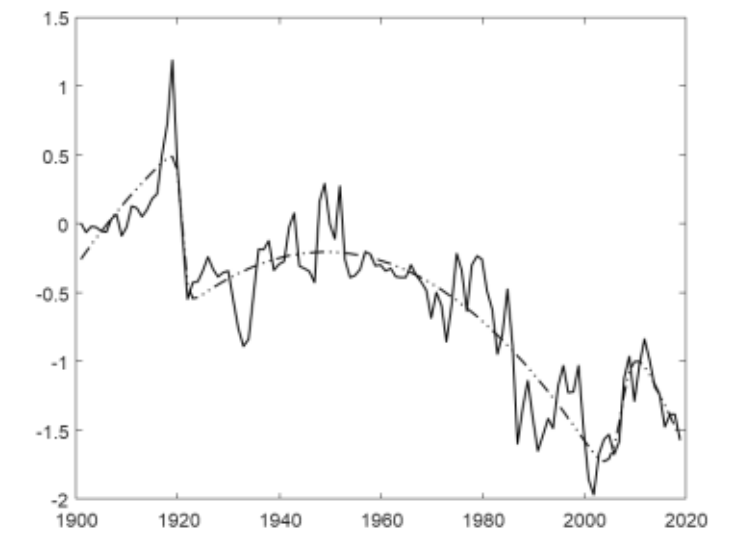 | 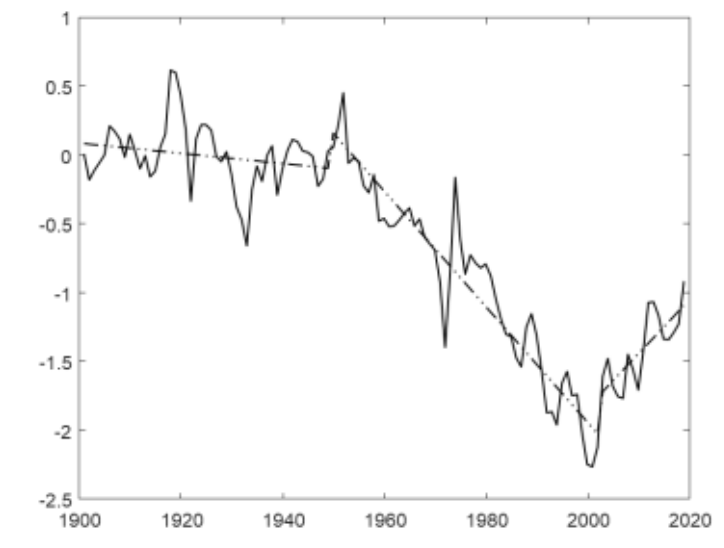 | 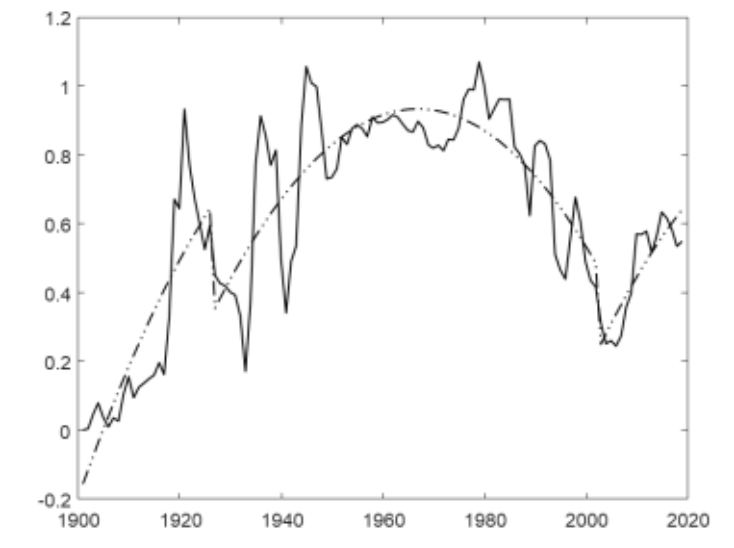 |

The prices of renewable commodities

**Table 1.** Estimation of the number of structural changes. Results for linear and quadratic models.

| | LINEAR SPECIFICATION | | | | | | | | QUADRATIC SPECIFICATION | | | | | | | |
|---|---|---|---|---|---|---|---|---|---|---|---|---|---|---|---|---|
| | *ExpW* (Model III) | *ExpW* (Model III; unrestricted) | *ExpW*(2/1) (Model III) | *ExpW*(2/1) (Model III; unrestricted) | No. of changes | *U* | No. of level changes | Model | *ExpW* (Model III) | *ExpW* (Model III; unrestricted) | *ExpW*(2/1) (Model III) | *ExpW*(2/1) (Model III; unrestricted) | No. of changes | *U* | No. of level changes | Model |
| Coffee | 1.741 | 0.093 | 2.124 | -0.214 | 0 | 0.359 | 0 | 0 | 1.750 | 0.103 | 2.853 | 0.787 | 0 | 0.353 | 0 | 0 |
| Cocoa | 4.421$^{b}$ | 0.067 | 1.179 | 0.088 | 1 | 0.398 | 0 | III(1) | 4.512$^{b}$ | 0.417 | 6.429$^{a}$ | 0.791 | 2 | 0.398 | 0 | III(2) |
| Tea | 3.586$^{a}$ | 0.008 | 8.678$^{b}$ | 0.040 | 2 | 0.306 | 0 | III(2) | 7.244$^{b}$ | 6.523$^{b}$ | 7.297$^{b}$ | -0.109 | 2 | | | III(2) |
| Rice | 10.652$^{b}$ | 0.035 | 12.185$^{b}$ | -0.051 | 2 | 0.359 | 0 | III(2) | 11.222$^{b}$ | 0.158 | 12.159$^{b}$ | 0.530 | 2 | 0.350 | 0 | III(2) |
| Wheat | 5.985$^{b}$ | 0.154 | 8.561$^{b}$ | 0.342 | 2 | 0.368 | 0 | III(2) | 6.018$^{b}$ | 0.376 | 9.962$^{b}$ | 1.496 | 2 | 0.366 | 0 | III(2) |
| Maize | 4.161$^{b}$ | 0.105 | 3.911 | -0.192 | 1 | 0.290 | 0 | III(1) | 4.202$^{a}$ | 0.280 | 4.260 | 0.306 | 1 | 0.312 | 0 | III(1) |
| Sugar | 2.855 | -0.038 | 9.519$^{b}$ | -0.070 | 0/2 | 0.445 | 0 | 0/III(2) | 6.890$^{b}$ | 0.064 | 27.038$^{b}$ | 18.393$^{b}$ | 2 | | | III(2) |
| Beef | 4.038$^{b}$ | 0.005 | 5.563$^{a}$ | -0.148 | 2 | 0.471 | 0 | III(2) | 4.071$^{a}$ | 0.251 | 41.629$^{b}$ | 41.477$^{b}$ | 2 | | | III(2) |
| Lamb | 2.846 | -0.024 | 4.941$^{a}$ | 0.163 | 0/2 | 0.477 | 0 | 0/III(2) | 3.006 | 0.188 | 5.036 | 4.244$^{b}$ | 0/2 | | | 0/III(2) |
| Banana | 7.905$^{b}$ | 0.317 | 10.544$^{b}$ | 0.147 | 2 | 0.296 | 0 | III(2) | 8.726$^{b}$ | 0.748 | 8.230$^{b}$ | 7.751$^{b}$ | 2 | | | III(2) |
| Palm oil | 3.345$^{a}$ | 0.099 | 4.124 | 0.331 | 1 | 0.385 | 0 | III(1) | 3.741$^{a}$ | 0.053 | 14.481$^{b}$ | 12.266$^{b}$ | 2 | | | III(2) |
| Cotton | 6.908$^{b}$ | 1.114 | 8.761$^{b}$ | -0.131 | 2 | 0.272 | 0 | III(2) | 6.587$^{b}$ | 0.147 | 10.475$^{b}$ | 1.040 | 2 | 0.260 | 0 | III(2) |
| Jute | 3.030 | 0.042 | 22.457$^{b}$ | 22.090$^{b}$ | 0/2 | | | 0/III(2) | 3.035 | 0.158 | 36.801$^{b}$ | 31.139$^{b}$ | 0/2 | | | 0/III(2) |
| Wool | 2.423 | 0.225 | 15.027$^{b}$ | 14.980$^{b}$ | 0/2 | | | 0/III(2) | 23.040$^{b}$ | 22.079$^{b}$ | 1.729 | -0.149 | 1 | | | III(1) |
| Hides | 3.263$^{a}$ | 0.115 | 4.720 | 4.190$^{b}$ | 2 | | | III(2) | 3.246 | 0.039 | 3.817 | 3.104$^{a}$ | | 0.668$^{b}$ | 1 | I(1) |
| Tobacco | 5.624$^{b}$ | 0.454 | 4.751 | 0.050 | 1 | 0.529$^{a}$ | 1 | I(1) | 5.394$^{b}$ | 0.256 | 16.555$^{b}$ | 16.078$^{b}$ | 2 | | | III(2) |
| Rubber | 6.279$^{b}$ | 0.514 | 16.009$^{b}$ | 13.350$^{b}$ | 2 | | | III(2) | 6.673$^{b}$ | 0.310 | 20.112$^{b}$ | 19.859$^{b}$ | 2 | | | III(2) |
| Timber | 3.880$^{a}$ | 0.168 | 6.122$^{b}$ | 0.177 | 2 | 0.437 | 0 | III(2) | 3.718$^{a}$ | 0.177 | 24.297$^{b}$ | 20.213$^{b}$ | 2 | | | III(2) |

*ExpW* refers to the statistic to test for the stability of the trend function. General and unrestricted cases; linear (Perron and Yabu, 2009) and quadratic (Presno *et al*., 2014) specifications. Trimming parameters: $\varepsilon = 0.05$, and $\delta = 0.5$.
*ExpW*(2/1) denotes the statistic to test the null of one versus two breaks; linear (Kejriwal and Perron, 2010) and quadratic (Presno *et al*., 2014) specifications. Trimming parameters: $\varepsilon = 0.1$; $\delta = 0.1$.
*U* is the statistic to detect breaks in level; linear (Harvey *et al.*, 2010) and quadratic (Presno *et al*., 2014) specifications. Trimming parameter: $m = 0.1$.
In case of contradiction between the tests, we opted for the more general model.
The column 'Model' indicates the model selected, with the number of changes in parentheses.
$^{a, b}$ denotes significance at 10% and 5%, respectively.

**Table 2.** Results of model selection criteria.

| | | BREAK LINEAR | SMOOTH LINEAR | BREAK QUADRATIC | SMOOTH QUADRATIC |
|---|---|---|---|---|---|
| No. changes | **Coffee** | 0 | 0 | **0** | 0 |
| SIC | | -175.87 | -175.87 | **-203.67** | -203.67 |
| AIC | | -181.43 | -181.43 | **-212.01** | -212.01 |
| Adj. $R^2$ | | 0.038 | 0.038 | **0.262** | 0.262 |
| No. changes | **Cocoa** | 1 | 1 | **2** | 2 |
| SIC | | -209.31 | -213.64 | **-265.74** | -256.01 |
| AIC | | -223.20 | -230.31 | **-290.76** | -286.58 |
| Adj. $R^2$ | | 0.543 | 0.573 | **0.749** | 0.744 |
| No. changes | **Tea** | 2 | 2 | 2 | **2** |
| SIC | | -384.12 | -395.52 | -404.21 | **-407.15** |
| AIC | | -406.35 | -423.31 | -429.22 | **-437.72** |
| Adj. $R^2$ | | 0.844 | 0.867 | 0.873 | **0.883** |
| No. changes | **Rice** | 2 | 2 | 2 | **2** |
| SIC | | -312.44 | -305.64 | -323.17 | **-316.55** |
| AIC | | -334.68 | -333.43 | -347.18 | **-348.12** |
| Adj. $R^2$ | | 0.845 | 0.845 | 0.862 | **0.863** |
| No. changes | **Wheat** | 2 | 2 | **2** | 2 |
| SIC | | -287.69 | -287.06 | **-292.34** | -284.16 |
| AIC | | -309.44 | -314.24 | **-316.81** | -314.06 |
| Adj. $R^2$ | | 0.746 | 0.760 | **0.764** | 0.762 |
| No. changes | **Maize** | 1 | **1** | 1 | 1 |
| SIC | | -290.28 | **-290.22** | -289.59 | -274.31 |
| AIC | | -304.17 | **-306.90** | -306.26 | -293.34 |
| Adj. $R^2$ | | 0.771 | **0.778** | 0.777 | 0.704 |
| No. changes | **Sugar** | 0 **2** | 0 2 | 2 | 2 |
| SIC | | -171.39 **-197.45** | -171.39 -190.61 | -193.17 | -186.53 |
| AIC | | -176.94 **-219.68** | -176.94 -218.40 | -218.18 | -217.10 |
| Adj. $R^2$ | | 0.469 **0.647** | 0.469 0.648 | 0.645 | 0.647 |
| No. changes | **Beef** | 2 | 2 | 2 | **2** |
| SIC | | -270.13 | -286.76 | -275.30 | **-289.31** |
| AIC | | -292.36 | -314.55 | -300.32 | **-319.88** |

| | | BREAK LINEAR | SMOOTH LINEAR | BREAK QUADRATIC | SMOOTH QUADRATIC |
|---|---|---|---|---|---|
| No. changes | **Banana** | 2 | 2 | 2 | **2** |
| SIC | | -504.20 | -499.80 | -501.20 | **-500.44** |
| AIC | | -526.43 | -527.59 | -526.21 | **-531.01** |
| Adj. $R^2$ | | 0.820 | 0.825 | 0.821 | **0.831** |
| No. changes | **Palm oil** | 1 | 1 | **2** | 2 |
| SIC | | -274.52 | -270.39 | **-306.23** | -299.96 |
| AIC | | -288.42 | -287.07 | **-331.24** | -330.53 |
| Adj. $R^2$ | | 0.767 | 0.766 | **0.843** | 0.844 |
| No. changes | **Cotton** | 2 | **2** | 2 | 2 |
| SIC | | -322.35 | **-326.39** | -324.76 | -322.82 |
| AIC | | -344.59 | **-354.18** | -349.78 | -353.39 |
| Adj. $R^2$ | | 0.884 | **0.894** | -0.890 | 0.894 |
| No. changes | **Jute** | 0 2 | 0 2 | 0 2 | 0 **2** |
| SIC | | -225.86 -305.48 | -225.86 -331.88 | -244.13 -333.28 | -244.13 **-330.86** |
| AIC | | -231.42 -327.72 | -231.42 -359.67 | -252.47 -358.29 | -252.47 **-361.43** |
| Adj. $R^2$ | | 0.419 0.753 | 0.419 0.814 | 0.517 0.811 | 0.517 **0.819** |
| No. changes | **Wool** | 0 **2** | 0 2 | 1 | 1 |
| SIC | | -228.27 **-331.04** | -228.27 -322.28 | -332.74 | -327.03 |
| AIC | | -233.83 **-353.27** | -233.83 -350.07 | -349.42 | -346.48 |
| Adj. $R^2$ | | 0.738 **0.908** | 0.738 0.907 | 0.904 | 0.902 |
| No. changes | **Hides** | **2** | 2 | 1 (in levels) | 1 (in levels) |
| SIC | | **-298.73** | -289.00 | -287.11 | -276.56 |
| AIC | | **-320.96** | -316.79 | -303.79 | -293.23 |
| Adj. $R^2$ | | **0.745** | 0.740 | 0.701 | 0.673 |
| No. changes | **Tobacco** | 1 (in levels) | 1 (in levels) | **2** | 2 |
| SIC | | -387.18 | -426.17 | **-442.83** | -435.27 |
| AIC | | -398.30 | -440.07 | **-467.84** | -465.84 |
| Adj. $R^2$ | | 0.621 | 0.735 | **0.797** | 0.797 |
| No. changes | **Rubber** | 2 | 2 | 2 | **2** |
| SIC | | -200.05 | -206.01 | -203.42 | **-207.38** |
| AIC | | -221.80 | -233.20 | -227.89 | **-237.29** |

| | | | | | | | | | | | | | | |
|---|---|---|---|---|---|---|---|---|---|---|---|---|---|---|
| Adj. $R^2$ | | 0.779 | | 0.820 | | 0.795 | | **0.829** | | | 0.874 | 0.888 | 0.882 | **0.893** |
| No. changes | **Lamb** | 0 | 2 | 0 | 2 | 0 | 2 | 0 | **2** | **Timber** | 2 | 2 | 2 | **2** |
| SIC | | -217.79 | -282.45 | -217.79 | -297.54 | -214.53 | -285.59 | -214.53 | **-295.60** | | -346.16 | -362.30 | -359.16 | **-368.80** |
| AIC | | -223.35 | -304.68 | -223.35 | -325.34 | -222.86 | -310.60 | -222.86 | **-326.17** | | -368.40 | -390.09 | -384.17 | **-399.37** |
| Adj. $R^2$ | | 0.596 | 0.806 | 0.596 | 0.839 | 0.598 | 0.816 | 0.598 | **0.841** | | 0.592 | 0.665 | 0.645 | **0.692** |

SIC = Schwartz's information criterion. AIC = Akaike's information criterion. Adj. $R^2$ = Adjusted R-squared. The model specification selected by the complexity penalization criteria appears in bold.

**Table 3**. Results of stationarity testing for the models chosen by the model selection criteria.

| | Model | *k*=0.5 | *k*=0.8 | *k*=0.9 | *c.v.* 10% | *c.v.* 5% | *c.v.* 1% |
|---|---|---|---|---|---|---|---|
| Coffee | Quadratic | 0.0796[a] | 0.0665 | 0.0666 | 0.0736 | 0.0869 | 0.1212 |
| Cocoa | Quadratic Break III(2) | 0.0431[b] | 0.0432[b] | 0.0432[b] | 0.0322 | 0.0356 | 0.0436 |
| Tea | Quadratic Smooth III(2) | 0.0215 | 0.0215 | 0.0215 | 0.0236 | 0.0266 | 0.0341 |
| Rice | Quadratic Smooth III(2) | 0.0277 | 0.0299 | 0.0299 | 0.0327 | 0.0376 | 0.0502 |
| Wheat | Quadratic Break III(2) | 0.0364 | 0.0385 | 0.0385 | 0.0407 | 0.0459 | 0.0589 |
| Maize | Linear Smooth III(1) | 0.0318 | 0.0329 | 0.0329 | 0.0508 | 0.0608 | 0.0855 |
| Sugar | Linear Break III(2) | 0.0328 | 0.0336 | 0.0336 | 0.0402 | 0.0452 | 0.0590 |
| Beef | Quadratic Smooth III(2) | 0.0317[c] | 0.0358[c] | 0.0358[c] | 0.0218 | 0.0246 | 0.0310 |
| Lamb | Quadratic Smooth III(2) | 0.0346[b] | 0.0358[b] | 0.0358[b] | 0.0249 | 0.0283 | 0.0364 |
| Banana | Quadratic Smooth III(2) | 0.0241[b] | 0.0257[c] | 0.0257[c] | 0.0179 | 0.0200 | 0.0248 |
| Palm oil | Quadratic Break III(2) | 0.0277 | 0.0291 | 0.0291 | 0.0332 | 0.0368 | 0.0469 |
| Cotton | Linear Smooth III(2) | 0.0312[b] | 0.0347[b] | 0.0347[b] | 0.0247 | 0.0278 | 0.0348 |
| Jute | Quadratic Smooth III(2) | 0.0288[b] | 0.0288[b] | 0.0288[b] | 0.0249 | 0.0281 | 0.0354 |
| Wool | Linear Break III(2) | 0.0431[a] | 0.0427[a] | 0.0427[a] | 0.0422 | 0.0483 | 0.0619 |
| Hides | Linear Break III(2) | 0.0791[b] | 0.0771[b] | 0.0771[b] | 0.0489 | 0.0565 | 0.0800 |
| Tobacco | Quadratic Break III(2) | 0.0342 | 0.0377[a] | 0.0377[a] | 0.0369 | 0.0427 | 0.0553 |
| Rubber | Quadratic Smooth III(2) | 0.0444[b] | 0.0455[b] | 0.0455[b] | 0.0362 | 0.0420 | 0.0561 |
| Timber | Quadratic Smooth III(2) | 0.0409[b] | 0.0460[b] | 0.0460[b] | 0.0342 | 0.0408 | 0.0549 |

In column 2, 'Quadratic' denotes the simple quadratic model (with no structural changes). In all the other cases, the name of the model indicates the nature of the model (linear/quadratic), the kind of structural changes (break/smooth transition), the specific model (I/III), and the number of structural changes detected (in parentheses).
*k* denotes a user-supplier constant (*k* = 0.5, 0.8, 0.9) required by the data-driven device employed to determine the bandwidth in the non-parametric estimator for the long-run variance of the process (Kurozumi, 2002).
*c.v.* denotes the critical value at 10%, 5%, and 1%, respectively.
[a, b, c] denotes significance at 10%, 5%, and 1%, respectively.

**Table 4.** Structural change dates and parameter estimates.

| | Model | Change dates | Parameter estimates |
|---|---|---|---|
| Coffee<br>(level) | Quadratic<br>No changes | - | |
| Cocoa<br>(differences) | Quadratic Break<br>III(2) | 1947<br>1973 | $\hat{\lambda}_1$=0.390<br>$\hat{\lambda}_2$=0.610 |
| Tea<br>(level) | Quadratic Smooth<br>III(2) | 1970 (midpoint)<br>1977 (midpoint) | $\hat{\lambda}_1$=0.585; $\hat{\gamma}_1$=19.838<br>$\hat{\lambda}_2$=0.647; $\hat{\gamma}_2$=228.634 |
| Rice<br>(level) | Quadratic Smooth<br>III(2) | 1931 (midpoint)<br>2007 (midpoint) | $\hat{\lambda}_1$=0.264; $\hat{\gamma}_1$=228.641<br>$\hat{\lambda}_2$=0.901; $\hat{\gamma}_2$=101.225 |
| Wheat<br>(level) | Quadratic Break<br>III(2) | 1930<br>2002 | $\hat{\lambda}_1$= 0.2679<br>$\hat{\lambda}_2$ = 0.9107 |
| Maize<br>(level) | Linear Smooth<br>III(1) | 1988 (midpoint) | $\hat{\lambda}_1$=0.743; $\hat{\gamma}_1$=31.339 |
| Sugar<br>(level) | Linear Break<br>III(2) | 1925<br>1982 | $\hat{\lambda}_1$=0.210<br>$\hat{\lambda}_2$=0.689 |
| Beef<br>(differences) | Quadratic Smooth<br>III(2) | 1976<br>1982 | $\hat{\lambda}_1$=0.633; $\hat{\gamma}_1$=288.082<br>$\hat{\lambda}_2$=0.689; $\hat{\gamma}_2$=163.320 |
| Lamb<br>(differences) | Quadratic Smooth<br>III(2) | 1916<br>1951 | $\hat{\lambda}_1$=0.128; $\hat{\gamma}_1$=288.082<br>$\hat{\lambda}_2$=0.422; $\hat{\gamma}_2$=170.077 |
| Banana<br>(differences) | Quadratic Smooth<br>III(2) | 1920<br>2003 | $\hat{\lambda}_1$=0.165; $\hat{\gamma}_1$=288.082<br>$\hat{\lambda}_2$=0.868; $\hat{\gamma}_2$=288.082 |
| Palm oil<br>(level) | Quadratic Break<br>III(2) | 1920<br>1986 | $\hat{\lambda}_1$=0.168<br>$\hat{\lambda}_2$=0.723 |
| Cotton<br>(differences) | Linear Smooth<br>III(2) | 1922<br>2011 | $\hat{\lambda}_1$=0.175; $\hat{\gamma}_1$=288.081<br>$\hat{\lambda}_2$=0.932; $\hat{\gamma}_2$=288.081 |
| Jute<br>(differences) | Quadratic Smooth<br>III(2) | 1922<br>1987 | $\hat{\lambda}_1$=0.175; $\hat{\gamma}_1$=288.082<br>$\hat{\lambda}_2$=0.727; $\hat{\gamma}_2$=288.082 |
| Wool<br>(level) | Linear Break<br>III(2) | 1949<br>2002 | $\hat{\lambda}_1$=0.412<br>$\hat{\lambda}_2$=0.857 |
| Hides<br>(differences) | Linear Break<br>III(2) | 1930<br>1933 | $\hat{\lambda}_1$=0.246<br>$\hat{\lambda}_2$=0.271 |
| Tobacco<br>(level) | Quadratic Break<br>III(2) | 1926<br>2002 | $\hat{\lambda}_1$=0.219<br>$\hat{\lambda}_2$=0.857 |
| Rubber<br>(differences) | Quadratic Smooth<br>III(2) | 1926<br>1934 | $\hat{\lambda}_1$=0.228; $\hat{\gamma}_1$=288.082<br>$\hat{\lambda}_2$=0.298; $\hat{\gamma}_2$=244.959 |
| Timber<br>(differences) | Quadratic Smooth<br>III(2) | 1918<br>1923 | $\hat{\lambda}_1$=0.142; $\hat{\gamma}_1$=150.603<br>$\hat{\lambda}_2$=0.184; $\hat{\gamma}_2$=288.081 |

'Level' indicates that the model has been fitted to the series in levels.
'Differences' indicates that the model has been fitted to the differenced series.
In models I and III with smooth (logistic sigmoid) transition functions, $\hat{\lambda}_j$and $\hat{\gamma}_j$ denote the non-linear least squares estimates for parameters $\lambda_j$ and $\gamma_j$ , respectively. In the break models, $\hat{\lambda}_j$ denotes the non-linear least squares estimate for $\lambda_j = \boldsymbol{\pi}_j$.